\documentclass[aps,pra,superscriptaddress,12pt,tightenlines,nofootinbib]{revtex4}
\usepackage{amsmath,amsthm,graphicx,amssymb,verbatim,listings}
\usepackage[T1]{fontenc}     
\usepackage{lmodern}         
\usepackage[ pdftex, plainpages = false, pdfpagelabels,
                 pdfpagelayout = useoutlines,
                 bookmarks,
                 bookmarksopen = true,
                 bookmarksnumbered = true,
                 breaklinks = true,
                 linktocpage=all,
                 pagebackref=false,
                 colorlinks = true,
                 linkcolor = BrickRed,
                 urlcolor  = blue,
                 citecolor = BrickRed,
                 anchorcolor = green,
                 hyperindex = true,
                 hyperfigures
                 ]{hyperref}
\usepackage[usenames, dvipsnames]{xcolor}
\usepackage{xifthen} 

\newcommand{\tr}{\hbox{tr}}
\newcommand{\moment}[1]{{\ensuremath{\left\langle #1 \right\rangle}}}
\newcommand{\ket}[1]{{\ensuremath{\left| #1 \right\rangle}}}
\newcommand{\bra}[1]{{\ensuremath{\left\langle #1 \right|}}}
\newcommand{\braket}[2]{{\ensuremath{\left\langle #1 \middle| #2
      \right\rangle}}}
\newcommand{\ketbra}[2]{{\ensuremath{\left| #1 \middle\rangle \middle\langle #2
      \right|}}}
\newcommand{\expgiv}[2]{{\ensuremath{\left\langle #1 : #2
      \right\rangle}}}
\newcommand{\rhoghz}{{\ensuremath{\rho_{\rm GHZ}}}}

\makeatletter
\DeclareRobustCommand{\qbar}{\mathord{%
  \text{$\m@th\mkern-2mu\raisebox{-1.6ex}[0pt][0pt]{$\mathchar'26$}\mkern-9mu q$}%
}}
\makeatother

\newcommand{\arxiv}[2][]{\ifthenelse{\isempty{#1}}{\href{http://arxiv.org/abs/#2}{{\tt arXiv:\allowbreak{}#2}}} {\href{http://arxiv.org/abs/#2}{{\tt arXiv:\allowbreak{}#2 [#1]}}}}

\newcommand{\cO}{\mathcal{O}}

\newcommand{\hvd}{\varrho}
\newcommand{\effset}{\mathcal{E}}

\begin{document}

\title{Is the SIC Outcome There When Nobody Looks?}

\author{Blake C.\ Stacey}
\affiliation{QBism Research Group, Physics Department,\\ University of Massachusetts Boston}

\date{\today}

\begin{abstract}
  Informationally complete measurements are a dramatic discovery of
  quantum information science, and the \emph{symmetric} IC
  measurements, known as SICs, are in many ways optimal among
  them. Close study of three of the ``sporadic SICs'' reveals an
  illuminating relation between different ways of quantifying the
  extent to which quantum theory deviates from classical expectations.
\end{abstract}

\maketitle

\begin{flushright}
Any headline ending in a question mark can be answered with the single word \emph{No.}

--- journalist's adage\\(the Davis--Hinchliffe--Betteridge Law)
\end{flushright} 

\section{Introduction}

What feature of quantum physics distinguishes it from classical
mechanics?  Schr\"odinger's answer was
``entanglement''~\cite{Schroedinger:1935}. Today, though, this
response is rather pass\'e. On the one hand, we have learned that
entanglement is not unique to quantum mechanics, but occurs rather
generically in nonclassical theories that lack superluminal
signalling~\cite{Barrett:2007, Barnum:2012}. On the other hand, we
know that the mere occurrence of entanglement in a theory is,
quantifiably, less exotic than the violation of a Bell
inequality~\cite{Werner:1989, Gottesman:1999, Spekkens:2007,
  Spekkens:2016b}.  And if our answer is ``quantum phenomena can
violate a Bell inequality'', then a new question naturally arises.  Is
the specific numerical extent to which quantum theory violates a Bell
inequality meaningful, and why?  This article addresses that question
by relating two measures of departure from classicality, one grounded
in a Bell inequality and the other in recent progress on representing
quantum theory in wholly probabilistic terms.

To accept quantum theory is to strive to maintain a certain peculiar
consistency among expectations for mutually exclusive experiments. For
example, consider the paradigmatic double-slit scenario.  When only
slit number 1 is open, the probability of a detector click as a
function of detector position is given by some function, call it
$P_1(x)$.  This function is quite mundane: It takes only nonnegative
values and behaves in all ways like a classical probability.  The same
is true for the curves $P_2(x)$ and $P_{12}(x)$.  The puzzle is that
$P_{12}(x) \neq P_1(x) + P_2(x)$.  Closer investigation reveals that the
mere fact of interference is not as enigmatic as it first appeared,
and a Bell test can interrogate the exotic character of quantum
physics more stringently than the double-slit experiment can.
Remarkably, the same theme holds true in the more stringent
inquiry: Given any specific experimental arrangement, the
probabilities we compute with quantum theory appear quite ordinary.
It is in the \emph{meshing} of expectations for \emph{different}
interventions into the course of nature that the fundamental enigma of
quantum physics manifests.

We will study this using examples in Hilbert spaces of decreasing
dimension: first eight (three qubits), then four (two qubits) and
finally three (a single qutrit).  We begin with Mermin's three-qubit
Bell inequality~\cite{Mermin:1990, Mermin:1990b, Mermin:1993,
  Mermin:2016}.  From there, we will turn to the Hoggar
SIC~\cite{Hoggar:1981, Hoggar:1998, Szymusiak:2015, Stacey:2016,
  Stacey:2016b}, an eight-dimensional structure that provides a common
meeting ground for two ways of discussing the nonclassicality of
quantum theory.  On the one hand, it furnishes a SIC representation of
eight-dimensional quantum state space~\cite{Wootters:2009}, and so it
exemplifies the nonclassical meshing of probability assignments
described by Fuchs and Schack~\cite{Fuchs:2013, Stacey:2016c,
  Fuchs:2016b, Fuchs:2016a, Fuchs:2017b, DeBrota:2018}.  On the other
hand, that same state space is what one requires for the GHZ
\emph{gedankenexperiment} and for Mermin's three-qubit Bell
inequality.  Considering the Hoggar SIC will be enough to answer the
title question in the negative; we will then explore additional
nuances by developing the theme using qubit and qutrit SICs.  (The
SICs for single qubits and qutrits, as well as the Hoggar SIC in
dimension 8, stand out in some ways from the other known SICs and are
collectively known as the \emph{sporadic SICs}~\cite{Stacey:2016,
  Appleby:2017b, Waldron:2018}.)  Finally, we will conclude with some
thoughts on the project of reconstructing quantum theory from physical
principles.

\section{Mermin's Three-Qubit Bell Inequality}

Let $X$, $Y$ and $Z$ denote the Pauli operators, and write $XXX$ for
$X \otimes X \otimes X$ and so forth.  Then we can write Mermin's
three-qubit Bell inequality~\cite{Mermin:1990b, Mermin:1993} in terms of
a linear combination of expectation values:
\begin{equation}
B(\rho) = \moment{XXX} - \moment{XYY} - \moment{YXY}
 - \moment{YYX}.
\label{eq:mermin-bell1}
\end{equation}
One employs this inequality in the following manner.  First, one
argues that the hypothesis of local hidden variables implies
\begin{equation}
-2 \leq B(\rho) \leq 2.
\label{eq:mermin-bell2}
\end{equation}
A way to see why these bounds should be set at $\pm 2$ is as follows.
Suppose that each part of the tripartite system carries a pair of
physical properties that respectively determine the outcomes of an $X$
measurement and of a $Y$ measurement performed on that part.  As a
whole, then, the system carries a set of properties
\begin{equation}
\lambda = (\lambda_{1X}, \lambda_{1Y}, \lambda_{2X}, \lambda_{2Y},
 \lambda_{3X}, \lambda_{3Y}),
\end{equation}
such that if we knew these values, we could say
\begin{equation}
\begin{array}{l}
 \moment{XXX} - \moment{XYY} - \moment{YXY}
 - \moment{YYX} \\
\qquad = \lambda_{X1}\lambda_{X2}\lambda_{X3}
   - \lambda_{X1}\lambda_{Y2}\lambda_{Y3}
   - \lambda_{Y1}\lambda_{X2}\lambda_{Y3}
   - \lambda_{Y1}\lambda_{Y2}\lambda_{X3}.
\end{array}
\end{equation}
It is now a matter of arithmetic to verify that for each assignment
of~$+1$ and $-1$ to the six $\lambda$s, this quantity is either $+2$
or $-2$.  The list of values denoted by $\lambda$ is, in older jargon, a
``dispersion-free state''~\cite{VonNeumann:1955, Werner:2014, Stacey:2016d,
  Mermin:2018}.  Since the sum of expectation values given any
dispersion-free state is $\pm 2$, any probabilistic average over
dispersion-free states will lie in the interval $[-2, 2]$.

Having established the bounds in (\ref{eq:mermin-bell2}), one then
finds a state --- for example, the GHZ state --- for which those bounds
are violated.  This establishes that quantum probabilities cannot be
accounted for by local hidden variables.  The GHZ state $\rhoghz$ is
(by definition~\cite{Mermin:1990}) an eigenstate of the operator $XXX$ with
eigenvalue $+1$, and it is also an eigenstate of~$XYY$, of~$YXY$ and
of~$YYX$ with eigenvalue $-1$.  Therefore, $B(\rhoghz) = 4$.  This
means that $\rhoghz$ violates the inequality (\ref{eq:mermin-bell2}),
and thus, the statistics encapsulated in~$\rhoghz$ defy local classical
emulation.

In discussions of hidden-variable models, it usually does not
particularly matter what other mathematical structure the set of all
$\lambda$'s might have.  The $\lambda$'s might, for all we end up
caring, be labeled by the elements of a group, or the morphisms of a
groupoid, or the open sets of a topology; they could have any
geometry, or none.  (Indeed, it is fair to say that the nature of
$\lambda$-space is ``rarely subject to much critical
scrutiny''~\cite{Fuchs:2014b}.)  What does matter is the hypothesis that
each part of a system carries its part of $\lambda$ with it as an
intrinsic physical property.  In the example above, we hypothesized
that each of the three qubits carried its own, intrinsic $\lambda_X$
and $\lambda_Y$.  A preparation of the system naturally corresponds,
then, to a probability distribution over the set of all possible
$\lambda$'s, or in other words, to a point in the simplex whose
vertices are labeled by the possible values of~$\lambda$.  We could
choose to decorate these vertices with additional structure (say,
making them into a group), but that extra mathematical ornamentation
is secondary to the \emph{physical} assumption which makes our state
space into a simplex and, ultimately, powers the derivation of Bell
inequalities~\cite{Werner:2014}.

\section{Probabilistic Representation of Quantum Theory}

A basic axiom of quantum theory is that to each physical system is
associated a complex Hilbert space. In the domain of quantum
information and computation, the Hilbert space is often taken to be
finite-dimensional, with the dimension $d$ scaling with the available
budget. We mathematically represent a measurement by a \emph{positive
  operator valued measure,} or POVM, which is a set of positive
semidefinite operators on the system's Hilbert space that sum to the
identity. Each operator in the set $\{E_i\}$ stands for an outcome of
the measurement. When a physicist --- let us call her Alice, per genre
tradition --- ascribes a quantum state to a system of interest, she
writes a positive semidefinite operator of unit trace, i.e., a density
matrix $\rho$. Alice's probability for obtaining the outcome $E_i$ is
\begin{equation}
  p(E_i) = \tr(\rho E_i).
  \label{eq:born}
\end{equation}
The POVM version of Gleason's theorem establishes that any assignment
of probabilities to experiment outcomes must take the form of
Eq.~(\ref{eq:born}) for some density matrix $\rho$, if the probability
of an outcome is independent of the POVM in which it is
embedded~\cite{Busch:2003, Caves:2004}.

If the elements of a POVM span the space of Hermitian operators, then
we can write any density matrix $\rho$ as a linear combination of the
POVM elements with real coefficients. This fact implies the
possibility of \emph{informationally complete} (IC) POVMs. Given a
probability vector $p$ over the outcomes of an IC POVM, we can
reconstruct the density matrix $\rho$, and so we can in principle do
anything we would have done with $\rho$ using $p$ instead. An IC POVM
must have at least $d^2$ elements to span the operator space. A
\emph{minimal} IC POVM, or MIC, has exactly $d^2$ elements. MICs can
be constructed in any dimension $d$~\cite{Caves:2002c, Appleby:2007a};
the question is how nice they can be made.

Let $\{\ket{\pi_i}\}$ be a set of $d^2$ unit vectors in a
$d$-dimensional complex Hilbert space that enjoy the following
property:
\begin{equation}
  |\braket{\pi_i}{\pi_j}|^2 = \frac{d\delta_{ij} + 1}{d+1}.
\end{equation}
Such a set is called a \emph{SIC,} where the \emph{S} stands for
``symmetric'' and the \emph{IC} for ``informationally complete'' as
before~\cite{Zauner:1999, Renes:2004, Scott:2010a, Fuchs:2017a}. The
rank-1 projection operators
\begin{equation}
  \Pi_i = \ketbra{\pi_i}{\pi_i}
\end{equation}
form, after an appropriate scaling, a MIC:
\begin{equation}
  H_i = \frac{1}{d}\Pi_i.
\end{equation}
Given a quantum state $\rho$, we have
\begin{equation}
  p(H_i) = \frac{1}{d}\tr(\rho\Pi_i),
\end{equation}
and we can reconstruct $\rho$ from these probabilities by way of an
appealingly simple formula:
\begin{equation}
  \rho = \sum_i \left[(d+1)p(H_i) - \frac{1}{d}\right]\Pi_i.
\label{eq:rho-from-p}
\end{equation}
Given any \emph{other} POVM $\{D_j\}$, we can find its outcome probabilities by
\begin{equation}
  q(D_j) = \tr(\rho D_j)
  = \sum_i\left[(d+1)p(H_i) - \frac{1}{d}\right] p(D_j|H_i),
  \label{eq:born2}
\end{equation}
where the conditional probability on the right-hand side is
\begin{equation}
  p(D_j|H_i) = \tr(D_j \Pi_i).
\end{equation}
Note that Eq.~(\ref{eq:born2}) has the form of the classical Law of
Total Probability
\begin{equation}
  p(D_j) = \sum_i p(H_i) p(D_j|H_i),
\end{equation}
but with the probabilities $p(H_i)$ ``deformed'' by a rescaling and a
shift. In prior work, the importance of Eq.~(\ref{eq:born2}) was
recognized by designating it the \emph{urgleichung} (``primal
equation'' in German, or perhaps Klingon).

Note that the bracketed quantity in the urgleichung, $(d+1)p(H_i) -
1/d$, can go negative if $p(H_i)$ is sufficiently small. This
deformation of the vector of $p(H_i)$ is technically what is sometimes
known as a ``quasi-probability'' --- a vector whose sum is normalized
to unity, but whose elements are not confined to the unit
interval~\cite{Feynman:1987, Ferrie:2008, Ferrie:2009, Zhu:2016a,
  DeBrota:2017}.  Negative ``quasi-probabilities'' are an artifact of
trying to squeeze something into the form of the Law of Total
Probability that doesn't actually fit. Generally, states that are
close to orthogonal to one of the SIC vectors will pick up negativity
in what we might call their quasi-probability representation. But it's
the $\{P(H_i)\}$ that are directly, operationally meaningful. There is
an experiment that Alice could go into the lab and do, and $P(H_i)$ is
how much she should bet on the $i$\textsuperscript{th} outcome of it.
Negativity of quasi-probability can become meaningful after one
introduces a notion of ``quasi-classical states'', or ``states that
are easy to emulate on a classical machine''. Once we bring in the
ideas necessary to support a ``resource theory'', then negativity can
gain significance as a measure of how powerful a given resource
is. But in the broader scheme of things, it is a secondary and
somewhat incidental notion.\footnote{Another reason to think of
  negativity as a secondary manifestation of nonclassicality is that
  we have considerable ``gauge freedom'' about where to put
  it. Adopting a vector notation for the urgleichung, we can express it as
  $Q(D) = P(D|H)\,\Phi\,P(H)$, where $\Phi$ is a linear combination of
  the identity and the all-ones matrix~\cite{DeBrota:2018}.  In this
  form, it is clear that we can multiply $\Phi$ to the right, turning
  the vector $P(H)$ into ``quasi-probabilities'', or we could multiply
  $\Phi$ to the left, putting the negativity into the conditional
  probability matrix $P(D|H)$. We could even express $\Phi$ as
  $\Phi^{1/2} \Phi^{1/2}$ and split the negativity across both.}

Any MIC will yield an expression akin to the urgleichung, but it can
be proven that out of all MICs, the SICs furnish the expression that
is as close as possible to the classical Law of Total
Probability~\cite{DeBrota:2018}. The intuition at work here is that,
classically, an informationally complete measurement would be, e.g.,
one that reads off a system's coordinates in phase space.  Any other
measurement would in principle be a coarse-graining of that
information.\footnote{Classical measurements can, of course, disturb
  the system that they are applied to.  But this is a largely
  uninteresting complication.  In order to express Pauli's ``ideal of
  the detached observer''~\cite{Gieser:2005, Atmanspacher:2009,
    Fuchs:2017b}, reading off the system's intrinsic properties
  without disruption is clearly the correct idealization.  Comparing
  the quantum and the classical in a reasonable way requires the
  plainest expression of both.}  But in quantum theory, there is no
underlying phase space, so we should not use a formula that depends
upon the concept of one.  By identifying this ``minimum distance''
between a probabilistic representation of quantum theory and classical
probability, SICs provide a measure of exactly how nonclassical
quantum physics is. This naturally raises the question of how this
measure of nonclassicality relates to other such, of which the
quintessential is the violation of a Bell inequality. We will answer
this question in the next section.

\section{The Hoggar SIC}
Consider the tensor product of three copies of qubit state space. We
will take for our computational basis the tensor-product basis of
Pauli $Z$ eigenstates.

Now, we construct the Hoggar SIC, which will provide a ``Bureau of
Standards'' experiment --- a reference measurement with respect to
which we can represent quantum theory in wholly probabilistic
terms~\cite{Appleby:2016b}. This construction is an example of how all
known SICs are generated: We begin with a \emph{fiducial vector} and
take its orbit under the action of a group~\cite{Renes:2004}. A
convenient fiducial for our present purpose is the vector given up to
normalization by
\begin{equation}
\ket{\pi_0^{\rm (Hoggar)}} \propto (-1+2i, 1, 1, 1,
                      1, 1, 1, 1)^{\rm T}.
\label{eq:hoggar-fiducial}
\end{equation}
We apply the three-qubit Pauli group to generate the Hoggar
SIC~\cite{Jedwab:2015, Szymusiak:2015, Stacey:2016}.  This is a set of
64 equiangular unit vectors $\{\ket{\pi_i}\}$, which we can also
represent in terms of the rank-1 projection operators $\Pi_i =
\ket{\pi_i}\bra{\pi_i}$. These define a representation of all
three-qubit states as probability vectors:
\begin{equation}
p(H_i) = \frac{1}{d} \tr(\rho \Pi_i).
\end{equation}
Given a probability distribution, we can construct the corresponding
density matrix using Eq.~(\ref{eq:rho-from-p}).\footnote{When we count
  SICs in a dimension $d$, we do so up to unitary equivalence, since
  an overall unitary transformation of the entire set preserves the
  inner products between vectors. There are 240 distinct SICs of
  ``Hoggar type''; i.e., by picking different fiducial vectors, one
  can construct 240 distinct sets of 64 lines apiece which all have
  the same symmetry and which are all orbits under the three-qubit
  Pauli group. All of these 240 sets are equivalent to one another
  under unitary or anti-unitary transformations~\cite{Zhu:2012}, so
  for brevity, we can refer to the SIC constructed from the fiducial
  (\ref{eq:hoggar-fiducial}) as \emph{the} Hoggar SIC.}

Note what happens if we take the expectation value of an operator:
\begin{equation}
\moment{A} = \tr(A \rho).
\end{equation}
Substituting in the expansion (\ref{eq:rho-from-p}), we obtain
\begin{align}
\moment{A} &= \tr \left[A \sum_i \left((d+1)p(H_i) - \frac{1}{d}\right)
                       \Pi_i\right] \\
 &= \sum_i \left((d+1) p(H_i) - \frac{1}{d}\right) \tr(A \Pi_i).
\end{align}
Denote the expectation value of an operator $A$ given the SIC state
$\Pi_i$ as
\begin{equation}
\expgiv{A}{i} = \tr(A \Pi_i).
\end{equation}
Then,
\begin{equation}
\moment{A} = (d+1) \sum_i p(H_i) \expgiv{A}{i}
 - \frac{1}{d}\sum_i \expgiv{A}{i}.
\end{equation}
We also know that
\begin{equation}
\sum_i \tr(A \Pi_i) = \tr\left[A \sum_i \Pi_i\right] = d\,\tr A.
\end{equation}

Each of the four operators $XXX$, $XYY$, $YXY$ and $YYX$ are
themselves traceless.  If we fix
\begin{equation}
\tr A = 0,
\end{equation}
then we obtain
\begin{equation}
\moment{A} = (d+1) \sum_i p(H_i) \expgiv{A}{i}.
\label{eq:boxed-expectation}
\end{equation}
This applies to each of the four operators, and also to linear
combinations of them.  It is more appropriate to use it for the
individual operators, since those correspond to individual
experiments, or to single trials in a multi-trial experiment.

If we followed classical intuition, we might say, ``The expectation
value for the random variable $A$, if the system is in configuration
$\Pi_i$, is some number $\expgiv{A}{i}$.  We don't know what
configuration the system is really in, so we have some probability
spread over~$i$.  To find the expectation value of~$A$, we just weight
the $\expgiv{A}{i}$ according to those probabilities.''  However, this
does not give the correct answer.  The classical result is off by a
factor $(d+1)$.

We can calculate the $\expgiv{A}{i}$ for the Hoggar SIC.  In fact, the
peculiar symmetry of the Hoggar SIC makes the salient features of the
computation rather easy to derive.  The four operators in Mermin's
inequality are elements of the group $\{D_k\}$ that generates the
Hoggar SIC.  Therefore, each of the four of them satisfies
\begin{equation}
\left|\bra{\psi_i} D_k \ket{\psi_i} \right|^2
 = \frac{1}{d+1} = \frac{1}{9}.
\end{equation}
Furthermore, each operator $D_k$ is Hermitian, so its eigenvalues are
real, as is its expectation value given any state.  Consequently,
\begin{equation}
\expgiv{D_k}{i} = \bra{\psi_i} D_k \ket{\psi_i} = \pm \frac{1}{3}.
\end{equation}
This applies to each term in our linear combination of expectation
values, Eq.~(\ref{eq:mermin-bell1}).  When we combine the expectation
values for the four operators, the contributions might cancel each
other, depending on the relative signs, but the absolute value of the
sum total cannot exceed $4/3$.  This is safely within the interval
that a local hidden variable explanation could account for.  So, the
Hoggar SIC states cannot be used as to violate the three-qubit Bell
inequality.  This will remain true for any SIC that is generated from
a fiducial by applying the three-qubit Pauli group.

By doing the algebra explicitly, we find that the Hoggar SIC states do
not even reach the bound of~$4/3$ that we deduced.  In fact,
\begin{equation}
|\expgiv{XXX}{i} - \expgiv{XYY}{i} - \expgiv{YXY}{i}
 - \expgiv{YYX}{i}|
 = \frac{2}{3}\ \forall\ i.
\end{equation}
Furthermore, any probabilistic combination of the Hoggar SIC states
will also be consistent with the LHV bound.  That is, if we pick a
state from the Hoggar SIC following the probability distribution
$p(H_i)$, then the linear combination of the four expectation values
will stay safely in the classical region.  If we then average
over~$i$, then this will remain true, no matter what the distribution
$p(H_i)$ is.

However!  The GHZ state itself corresponds to some probability
distribution $p_{\rm GHZ}(H_i)$, because we can write any state in the Hoggar SIC
representation.  Let the index $\cO$ range over the four
operators that we use to define the three-qubit Bell inequality:
\begin{equation}
\cO \in \{XXX, -XYY, -YXY, -YYX\}.
\end{equation}
For any of our four operators $\cO$,
\begin{equation}
  \sum_i p_{\rm GHZ}(H_i) \expgiv{\cO}{i} = \frac{1}{9},
\end{equation}
meaning that the quantum expectation value is scaled up by the
urgleichung's factor $d+1$:
\begin{equation}
  \moment{O} = (d+1)\sum_i p_{\rm GHZ}(H_i) \expgiv{\cO}{i} = 1.
\end{equation}
Therefore, 
\begin{equation}
\sum_{\cO} \sum_i p_{\rm GHZ}(H_i) \expgiv{\cO}{i} = \frac{4}{9}.
\end{equation}
This is within the classical interval $[-2,2]$, but when we account
for the extra factor in the urgleichung, we find
\begin{equation}
(d+1)\sum_{\cO} \sum_i p_{\rm GHZ}(H_i) \expgiv{\cO}{i} = 4.
\end{equation}
It is that factor of $(d+1)$ that lifts us over the edge into
nonclassical territory.

One way to interpret this result is as a bridge between
\emph{interference experiments} and \emph{Bell--Kochen--Specker
  phenomena.}  Interference phenomena are weakly nonclassical: That
is, the \emph{bare fact} of interference can occur in fundamentally
classical theories~\cite{Spekkens:2007, Spekkens:2016}.  However,
by adopting the proper mindset, we can strengthen the double-slit
experiment into a genuine test for nonclassicality.

Interference between \emph{nonorthogonal alternatives} --- in other
words, between alternative paths represented by nonorthogonal quantum
states --- can be a stronger test of nonclassicality than the
double-slit experiment as it is normally described.  This is because
generalizing to nonorthogonal states allows the ``which-way''
information to be the outcome of an informationally complete
measurement.  (Heuristically speaking, this ties in with the idea that
pre- and post-selection effects with nonorthogonal states are more
strongly nonclassical than they are when one considers only orthogonal
states~\cite{Spekkens:2016, Pusey:2015}.)

Mermin wrote that the $n$-qubit GHZ state ``combines two of the most
peculiar features of the quantum theory''~\cite{Mermin:1990b},
interference of probabilities and the failure of local hidden-variable
explanations.  Using the Hoggar SIC, we have found a concise
expression of this when $n = 3$.  Correlations that violate the
three-qubit Bell inequality encode a kind of interference that defies
mimicking by classical randomness.

Mermin's three-qubit Bell inequality is closely related to the GHZ
thought-experiment, which is sometimes touted as an example where the
distinction between quantum and classical is ``all-or-nothing''.  The
hypothesis of local, intrinsic hidden variables implies one result
with certainty, and quantum mechanics implies another, also with
certainty.  Stated carelessly, this can create the impression that
probabilities are not involved.  But a prediction made with
probability 0 or 1 is still a probabilistic statement.  Moreover, we
can see the nontrivial probabilities churning just below the surface.

In the GHZ scenario, Alice measures the $X$ observable on each of her
three qubits and checks the parity of the answer.  Writing $\ket{+}$
and $\ket{-}$ for the eigenstates of~$X$, and denoting the SIC
representation of the state $\ket{+++}$ by $p_{+++}$, she
calculates that
\begin{equation}
  \sum_i p_{\rm GHZ}(H_i) p_{+++}(H_i) = \frac{5}{288}.
\end{equation}
The same result holds for the other states of the same parity,
$p_{+--}$, $p_{-+-}$ and $p_{--+}$.  Classical intuition would lead
her to say that this number is the probability for obtaining each of
the odd-parity outcomes, given a preparation described by $p_{\rm
  GHZ}$.  In turn, the probability for getting \emph{any} odd-parity
outcome would be the sum of the probabilities for the four
alternatives. But she knows to take the quantum correction, which is given
by the urgleichung:
\begin{align}
  P({\rm odd}) = &\ d(d+1)\sum_i p_{\rm GHZ}(H_i)
         [p_{+++}(H_i) + p_{+--}(H_i) + p_{-+-}(H_i) + p_{--+}(H_i)] \nonumber\\
         &- \sum_i[p_{+++}(H_i) + p_{+--}(H_i) + p_{-+-}(H_i) + p_{--+}(H_i)].
\end{align}
This evaluates to
\begin{equation}
  P({\rm odd}) = 72 \cdot 4 \cdot \frac{5}{288} - 4 = 1.
\end{equation}
So, while ascribing the GHZ state does imply predictions with
probability unity, that unity arises from the combination of many
fractions.

\section{Qubit Pairs and Twinned Tetrahedral SICs}

In this section, we change perspective slightly. Instead of applying
one SIC measurement to the entirety of a tripartite system, we start
with a smaller SIC and apply measurements based on it to each of two
halves of a bipartite system. The end result will be a sharpened
intuition for the nonclassicality of qubit pairs.

We have seen how attempting to interpret a SIC outcome as a specific,
pre-existing physical property leads to a contradiction with the
predictions of quantum theory.  Any assumption which would incline us to
interpret SIC outcomes in this way is, therefore, an assumption that
would lead the unwitting physicist into error and would stand in the
way of using quantum theory fruitfully.  We can identify one such
counterproductive idea --- the \emph{EPR criterion of reality}~\cite{Einstein:1935}:
\begin{quotation}
\noindent If, without in any way disturbing a system one can [gather
  the information required to] predict with certainty (i.e., with
probability equal to unity) the value of a physical quantity, then
there exists an element of physical reality corresponding to this
physical quantity.
\end{quotation}
We now present a scenario in which the EPR criterion leads the
unwitting physicist to conclude that SIC outcomes are pre-existing,
specific ``elements of physical reality.''

Alice arranges the following experiment.  A device produces pairs of
qubits to which Alice ascribes a maximally entangled state.  Each
qubit then travels to one of two widely separated instruments, which
we can designate the left detector and the right detector.  The
detectors each have a control knob that can be turned to four
different settings.  Alice models the detectors using binary POVMs
defined using the states comprising two SICs.  The first SIC, which we
can denote $\{\Pi_i^+\}$, is a set of four projectors that together
form the vertices of a tetrahedron inscribed in the Bloch sphere.  The
second SIC, $\{\Pi_i^-\}$, forms the tetrahedron whose vertices are
antipodal to those of the first.  Together, the two tetrahedra form a
stellated octahedron.  When the knob on a detector is set to position
$i$, it implements the POVM
\begin{equation}
\{\Pi_i^+, I - \Pi_i^+\} = \{\Pi_i^+, \Pi_i^-\}.
\end{equation}

Consider first the case when Alice sets the two control knobs to the
same position.  She performs the measurement with one detector, say
the one on the left.  If she experiences the $+$ outcome, she can
predict with 100\% certainty that she would experience the $-$ outcome,
if she were to walk over to the right-hand detector and test the other
qubit.  Likewise, if she experiences the $-$ outcome on the left, she
can predict with a probability of unity that she will experience $+$
upon using the detector on the right.  This holds true for all four
values of the control setting $i$.

Alice, deciding to entertain the EPR criterion, concludes that there
exists within both particles emitted from the common source an
``element of physical reality'' that implies the outcome of each of
the binary tests.

What happens when Alice chooses to set the two detectors differently?
Now, if she performs test $i$ on the left and obtains the $+$ outcome,
she updates her state for the right-hand particle to~$\Pi_i^-$.  She
then performs the test for some detector setting $j \neq i$ on the
right.  Her probability of obtaining the $-$ outcome on the right is
\begin{equation}
\tr(\Pi_i^- \Pi_j^-) = \frac{1}{d+1} = \frac{1}{3}.
\end{equation}
Likewise, if Alice first experiences the $-$ outcome on the left, she
updates her state for the right-hand side to~$\Pi_i^+$, and her
probability for obtaining the $+$ result on the right is
\begin{equation}
\tr(\Pi_i^+ \Pi_j^+) = \frac{1}{3}.
\end{equation}
In summary, when Alice sets the detector controls to the same
position, her probability of an anti-coincidence ($+$ on one device,
$-$ on the other) is unity.  If she sets the detector controls
differently, her probability of anti-coincidence is $1/3$.

Can Alice account for these results in terms of hidden variables?
Guided by the EPR criterion, she postulates that each particle carries
an ``instruction set''~\cite{Mermin:1985} of the form $\lambda_0
\lambda_1 \lambda_2 \lambda_3$.  Each $\lambda_i$ is a pre-existing
physical property of some kind, intrinsic to a particle, which can be
thought of as taking values in the set $\{+, -\}$.  The value
of~$\lambda_i$ specifies the outcome of testing that particle with a
detector configured to setting $i$.

Alice hypothesizes that the source produces particle pairs with
anticorrelated instruction sets:
\begin{equation}
\left\{\begin{array}{c}
       ++-- \\ --++ \\ -++- \\ +--+ \\ +-+- \\ -+-+
       \end{array}
\right\}
\hbox{ with }
\left\{\begin{array}{c}
       --++ \\ ++-- \\ +--+ \\ -++- \\ -+-+ \\ +-+-
       \end{array}
\right\}.
\label{eq:six-instruction-sets}
\end{equation}
Alice finds that whichever instruction sets the particles carry, if
she configures her two detectors identically, these instruction sets
imply perfect anti-coincidence.  If she instead sets her detector
knobs to different positions, each choice of detector configurations
will produce anti-coincidence with probability $1/3$, provided that
all six of these instruction-set pairs occur with equal probability.

How should Alice proceed from this point?  She supposes, as a
physicist naturally would, that whatever an instruction set is, a
particle can carry one all by itself.  The source in this experiment,
Alice figures, happens to produce particles in pairs with perfectly
anti-correlated instruction sets.  To imagine that a particle
\emph{only} has an instruction set when it is produced as half of a
pair strikes her as a touch pathological.  A spinning top has angular
momentum whether or not it is started into motion at the same time as
another top, spun in the opposite direction.

Let $T(i)$ be Alice's probability for obtaining the $+$ outcome when
performing test $i$ on an isolated system.  Quantum theory tells us
that we can craft another measurement corresponding to the
four-outcome POVM
\begin{equation}
\left\{\frac{1}{2} \Pi_0,
  \frac{1}{2} \Pi_1,
  \frac{1}{2} \Pi_2,  
  \frac{1}{2} \Pi_3\right\}.
\label{eq:tetra-SIC}
\end{equation}
Alice's probability for obtaining outcome $i$ in this experiment is
\begin{equation}
p(H_i) = \frac{1}{d}\tr(\rho \Pi_i) = \frac{1}{2} T(i).
\end{equation}

What is Alice's interpretation of this four-outcome experiment in
terms of her hidden-variable hypothesis?  Suppose that she has $T(0) =
1$.  In quantum language, this means that her state for the system is
the projector $\Pi_0^+$.  Referring back to the instruction sets
listed in Eq.~(\ref{eq:six-instruction-sets}), Alice notes that three
of them predict $+$ for the binary test on the first element:
\begin{equation}
\{ ++--, +--+, +-+- \}.
\end{equation}
Selecting a $+$ at random from this list, Alice finds that she obtains
a $+$ in position $0$ with probability $1/2$, and in each of the other
positions with probability $1/6$.  So, she can interpret $p(H_i)$ as the
probability that a $+$ sign, chosen at random from all the $+$ signs
occurring in all possible instruction sets, falls in position $i$.

The hypothesis of instruction sets implies that the outcome of a
tetrahedral SIC measurement, Eq.~(\ref{eq:tetra-SIC}), is a classical
random variable.  To adapt Einstein's phrase, the SIC outcome is there
even when nobody looks.  Knowing that the SIC measurement is
informationally complete, and seeing that its outcome probabilities
are determined by the probability distribution over the six
instruction sets, we conclude that the distribution over the
instruction sets is all that is necessary to calculate the outcome
statistics for \emph{any} experiment.

There is another route to the instruction sets in
Eq.~(\ref{eq:six-instruction-sets}), which begins with a set of
desiderata that Spekkens provides for a noncontextual hidden-variable
model~\cite{Spekkens:2014}.  The guiding philosophy of the Spekkens
criteria is that two quantities which imply the same statistics should
have the same representation in terms of probability distributions
over the underlying hidden variables. If two preparations of a system
yield the same statistics for all possible measurements, then those
two preparations correspond to the same distribution over
$\lambda$. Likewise, if two measurements have the same statistics for
all possible preparations, then those two measurements correspond to
the same conditional probabilities of outcomes given $\lambda$'s. The
key quantities are effects, that is, positive semidefinite operators
that satisfy
\begin{equation}
0 < E \leq I.
\end{equation}
Call the set of all effects $\effset$.  By hypothesis, if Alice knows
the ontic state $\lambda$, she has a map from effects to probabilities:
\begin{equation}
  w: \effset \to [0,1].
\end{equation}
The function $w$ will generally depend upon $\lambda$.  What
properties will it satisfy?  First, it obeys a sum rule. For any
discrete set of effects $\{E_i\} \subset \effset$, if $\sum_i E_i$ is
also an effect, then
\begin{equation}
w\left(\sum_i E_i\right) = \sum_i w(E_i).
\end{equation}
We will only need the particular special case of this in which the sum
of the $\{E_i\}$ is the identity operator, i.e., when the set of
effects is a POVM. This is equivalent to saying that whatever the
underlying ontic state of the system, when Alice applies a
measurement, she is sure that \emph{something} has to happen.

Furthermore, for any effect $E \in \effset$ and real number $s \in
[0,1]$, if $sE \in \effset$, then
\begin{equation}
w(sE) = s w(E).
\end{equation}
Again, we will only need a special case of this, specifically the case
when $s = 1/2$. This is equivalent to saying that for any measurement,
we can post-process the outcome by flipping a fair coin.

The identity effect is assigned unit probability:
\begin{equation}
w(I) = 1.
\end{equation}
If Alice doesn't care at all about what she does, then her probability
of ``whatever'' happening is 1, regardless of the ontic state. When
else can she have certainty? If and only if the effect in
question is a projection~\cite{Spekkens:2005}:
\begin{equation}
w(E) \in \{0,1\} \hbox{ if and only if } E^2 = E.
\end{equation}

What do these conditions imply for a qubit SIC?  First, the SIC states
form a POVM when scaled down by the dimension:
\begin{equation}
\sum_i \frac{1}{2} \Pi_i = I.
\end{equation}
Therefore, it must be the case that
\begin{equation}
\sum_i w\left(\frac{1}{2}\Pi_i\right) = 1.
\end{equation}
In turn, by the post-processing assumption,
\begin{equation}
\sum_i w\left(\frac{1}{2}\Pi_i\right)
 = \frac{1}{2} \sum_i w(\Pi_i).
\end{equation}
Each $\Pi_i$ is a projector, so each $w(\Pi_i)$ on the right-hand side must
be either 0 or 1.  Because the sum total must be normalized, exactly
two terms are 0, while the other two both equal 1.  Consequently, the
instruction sets in Eq.~(\ref{eq:six-instruction-sets}) are the only
configurations of hidden variables that are compatible with
noncontextuality and with the structure of a qubit SIC measurement.

If we postulate that a tetrahedral SIC measurement $\{\frac{1}{2}
\Pi_i^+\}$ is possible, and we
assert that the hidden-variable description of the qubit is
noncontextual, then any quantum state for the qubit implies a
probability distribution $\hvd(\vec{\lambda})$ over the six instruction
sets in Eq.~(\ref{eq:six-instruction-sets}).  In turn, such a
probability distribution implies a $p(H_i)$, specified by
\begin{equation}
p(H_i) = \frac{1}{2} \sum_{\vec{\lambda}} \hvd(\vec{\lambda})
\delta_{\lambda_i,+}.
\end{equation}
This has a ready interpretation in terms of a two-step stochastic
process.  Effectively, we are picking an instruction set at random
with probability $\hvd(\vec{\lambda})$, and then we are flipping a
fair coin to select one of the two $+$ signs in that instruction set.

By mapping points in the Bloch ball to density operators, and then
solving for the corresponding hidden-variable distributions, we can
map out the ``classical region'' of qubit state space.  We define this
region to be the subset of state space within which all elements
of~$\hvd$ turn out nonnegative, meaning that the vector $\hvd$ can be
interpreted as an ordinary probability distribution, rather than a
quasiprobability one.  The eight states that comprise the vertices of
the SICs $\{\Pi_i^+\}$ and $\{\Pi_i^-\}$ are classical, by this
standard.  The classical region of state space is the cube that is
their convex hull.

Each of the six instruction sets in
Eq.~(\ref{eq:six-instruction-sets}) is a ``dispersion-free
state''~\cite{VonNeumann:1955, Werner:2014, Stacey:2016d,
  Mermin:2018}.  Using the SIC representation of qubit state space, we
can see that they do not correspond to valid quantum states.  Each
instruction set implies a probability distribution $p$ in which two
elements equal $1/2$ and the other two equal $0$.  Using
Eq.~(\ref{eq:rho-from-p}), we can map these probability distributions
to linear operators.  The resulting operators will all be Hermitian,
but they \emph{will not} be positive semidefinite.  Therefore, the
dispersion-free states cannot be quantum states.  Pictorially,
they can be represented as the vertices of an octahedron
\emph{outside} the Bloch sphere: While the Bloch sphere has radius 1,
the dispersion-free states all reside at a distance of~$\sqrt{3}$
from the origin.

We have seen that if we try to model the SIC states as essentially
classical, then the eigenstates of the Pauli operators become
maximally quantum, in that they lie as far as possible from the region
of the Bloch ball for which a classical model exists.  This is in a
sense the dual of the statement that qubit SIC states are ``magic
states'' when the Pauli eigenstates are treated as
classical~\cite{Bravyi:2005}.  Consequently, we now have a certain
intuition for the result of Andersson \emph{et al.,} who find that the
maximal violation in an ``elegant'' two-qubit Bell inequality occurs
when the measurements on one qubit are the Pauli eigenbases and the
measurements on the other are the binary tests defined by pairs of
antipodal SIC vectors~\cite{Andersson:2017}.

\section{Failure of Hidden Variables for Qutrits}

The Spekkens criteria for hidden-variable models provide an
alternative perspective on a Kochen--Specker proof that Bengtsson,
Blanchfield and Cabello derive for qutrit
systems~\cite{Bengtsson:2012}.  Their proof relies upon a set of 21
vectors, 9 of which comprise a SIC and the other 12 of which form a
particular set of orthonormal bases. These four bases have the nice
property that they are all \emph{mutually unbiased} with respect to
one another.  That is, the overlap of any vector from one basis with
any vector from another basis is constant. In the Bengtsson \emph{et
  al.}\ construction, the only properties that matter are the
orthogonalities among the 21 vectors; by employing the fact that the
set of 9 specifically form an informationally complete POVM, we can
appreciate the result in a new way.

First, we note that if we have three vectors that form an orthonormal
basis for~$\mathbb{C}^3$, then the projectors onto those vectors add to
the identity operator, meaning that by the Spekkens rules,
\begin{equation}
  w(E_1) + w(E_2) + w(E_3) = 1.
\end{equation}
Furthermore, each term in the sum must be either 0 or 1, implying that
whatever the underlying ontic state, exactly one vector in any
orthonormal basis is assigned probability~1.

Now, we consider the cat's cradle of vectors we encounter in dimension
$d = 3$.  First, there's the Hesse SIC.  Take $\omega=e^{2\pi i/3}$,
and construct the set of states $\{|\pi_j\rangle\}$ given by the
columns of
\begin{equation}
\frac{1}{\sqrt{2}}
\left(
\begin{array}{ccccccccc}
0 & -1 & 1 & 0 & -1 & 1 & 0 & -1 & 1 \\
1 & 0 & -1 & \omega & 0 & -\omega & \omega^2 & 0
 & -\omega^2 \\
-1 & 1 & 0 & -\omega^2 & \omega^2 & 0
 & -\omega & \omega & 0
\end{array}
\right).
\label{eq:ME-SIC}
\end{equation}
We have a duality relation between the canonical mutually unbiased
bases and the Hesse SIC.  This relation is rather intricate: Each of
the 9 SIC states is orthogonal to exactly 4 of the MUB states, and
each of the MUB states is orthogonal to exactly 3 SIC
states~\cite{Stacey:2016c}.

An easy way to remember these relationships is to consider the finite
affine plane on nine points.  Each of the points corresponds to a SIC
vector, and each of the lines correponds to a MUB vector, with
point-line incidence implying orthogonality.  The four bases are the
four ways of carving up the plane into parallel lines (horizontals,
verticals, diagonals and other diagonals).

Let the projectors onto the 9 SIC vectors be $\Pi_1$ through $\Pi_9$.
We can uniquely identify each of the projectors onto the MUB vectors
by the three SIC vectors to which they are
orthogonal~\cite{Stacey:2016c}.  For example, $M_{123}$ is orthogonal
to $\Pi_1$, $\Pi_2$ and $\Pi_3$.  The 12 MUB states are then
\begin{equation}
  \begin{array}{ccc}
    M_{123}, & M_{456}, & M_{789}; \\
    M_{147}, & M_{258}, & M_{369}; \\
    M_{159}, & M_{267}, & M_{348}; \\
    M_{168}, & M_{249}, & M_{357};
  \end{array}
\end{equation}
where each row corresponds to an orthonormal basis of $\mathbb{C}^3$.

Because $\Pi_1$ is orthogonal to $M_{123}$, if the underlying ontic
state $\lambda$ implies $w(M_{123}) = 1$, then $w(\Pi_1) = 0$.  The
more of the $\{w(\Pi_i)\}$ that we can ``zero out'' in this way, the
smaller their sum will be.  By working through all the possibilities
for assigning a $w$ of unity to exactly one element of each basis, it
is straightforward to show that whatever $\lambda$ might be,
\begin{equation}
  \sum_i w(\Pi_i) \leq 2.
\end{equation}
But from the post-processing rule,
\begin{equation}
  w\left(\frac{1}{d}\Pi_i\right) = \frac{1}{d} w(\Pi_i),
\end{equation}
and from the sum rule,
\begin{equation}
  \sum_i w\left(\frac{1}{d}\Pi_i\right)
  = w\left(\sum_i \frac{1}{d}\Pi_i\right)
  = w(I) = 1.
\end{equation}
Therefore,
\begin{equation}
  \sum_i w(\Pi_i) = d\sum_i w\left(\frac{1}{d}\Pi_i\right) = 3.
\end{equation}
Our plans for a hidden-variable model have gone awry.  The SIC states
burst out of the confines that the orthonormal bases establish.  The
set of 9 and the set of 12 cannot coexist in the world of~$\lambda$:
If we take one set to have a classical representation, then the other
cannot.

Bengtsson \emph{et al.}\ derive a contradiction between the hypothesis
of intrinsic hidden variables and the predictions of quantum theory by
invoking the Born rule. This is equivalent to postulating a
probability assignment for all POVMs, i.e., a frame function.  If we
know that the state space is the space of unit-trace positive
semidefinite operators and that probabilities are calculated by the
inner product between density matrices and effects, then we can say
that
\begin{equation}
  \sum_i \moment{\Pi_i} = \sum_i \tr(\rho\Pi_i) = d\,\tr\rho = 3.
\end{equation}
By invoking Spekkens' criteria for a hidden-variable model, we see
that we do not have to postulate outcome statistics for all POVMs.
Instead, we can derive the desired contradiction by considering only a
discrete set of rays in the Hilbert space $\mathbb{C}^3$.

\section{Quantum Theory from Nonclassical Probability Meshing}

Let us now return to the Hoggar SIC. We have seen how this
configuration, and the representation of quantum state space that it
furnishes, provides a link between \emph{interference phenomena} and
the \emph{failure of hidden-variable models.}  This mathematical
construction --- just sixty-four complex lines, making equal angles
with one another --- evidently cuts quite deeply into the quantum
mysteries.  Consider again our expression for the expectation value of
an operator:
\begin{equation}
\moment{A} = (d+1) \sum_i p(H_i) \expgiv{A}{i}
 - \frac{1}{d}\sum_i \expgiv{A}{i}.
\end{equation}
Seen in one way, this formula is a way to ``do Feynman right'': Like
the double-slit experiment, it captures the counterintuitive way
quantum theory requires us to use expectations for one scenario to
make deductions about another.  However, it indicates a kind of
interference that cannot be emulated by classical stochasticity.  And,
seen in another way, it opens the possibility of violating a Bell
inequality.

This is a sufficiently appealing notion that one is naturally tempted
to wonder how far it can go.  If we take this idea as basic, if we
make this way of relating expecations between counterfactual scenarios
as a fundamental precept, what can we derive from it?  The answer,
potentially, is \emph{quantum theory itself.}

During the twenty-first century, there has been increasing interest in
the project of \emph{rederiving} or \emph{reconstructing} the
mathematical apparatus of quantum mechanics, by starting with a set of
basic principles that, one hopes, are more meaningful or illuminating
than the abstruse invocations with which one traditionally begins the
subject~\cite{Fuchs:2016b}.  These efforts begin with a set of axioms,
typically expressed in operationalist terms as statements about what
kinds of laboratory procedures are possible, and rederive quantum
theory from that starting point~\cite{Barnum:2014, Barnum:2015,
  Wilce:2016, Krumm:2016, Cabello:2018,
  VanDeWetering:2018}. Mathematically, these derivations are
successful; however, they share the common feature that they make
quantum theory as ``benignly humdrum'' as
possible~\cite{Fuchs:2016b}. The remarkable and enigmatic phenomena
seen within quantum physics are no closer to the surface than they
were in the standard presentation of the
formalism~\cite{Appleby:2016b}. Indeed, the notions invoked in these
axioms are often not that quantum at all. For example, the system of
Chiribella, D'Ariano and Perinotti relies upon the purifiability of
mixed states~\cite{Chiribella:2011}, which was originally discovered
in quantum physics but actually arises naturally in the Spekkens toy
model, a fundamentally classical theory~\cite{Spekkens:2007,
  Disilvestro:2017}. Moreover, it is not at first glance clear what
these proposed sets of axioms have in common with each other, other
than their conclusion.

In contrast, one research program aims to take the urgleichung as a
basic postulate upon which quantum theory can be
built~\cite{Fuchs:2013, Fuchs:2016b, Appleby:2016b}.  The urgleichung
embodies a rejection of the hypothesis of hidden variables, phrased in
a way that does not depend upon the ordinary textbook formalism of
Hilbert spaces and operators.  The hope is that whatever deep lesson
quantum physics has to teach us about the character of the natural
world, we will see it most clearly by bringing the essential
expression of it to the forefront, rather than deriving it as a
consequence of ``benignly humdrum'' axioms.  Postulating that the
urgleichung is fundamental --- that, try as one might to establish a
standard reference measurement, nonclassical ``deformation'' of
probabilities \emph{cannot be avoided} --- foregrounds the strangeness
of quantum theory.  What we know so far is that a reconstruction on
these lines \emph{can} be done, but at the price of invoking a couple
additional presumptions that seem too specific to belong in the final
answer.  My own suspicion is that these additional requirements are
stronger and more particular than is truly necessary.  This is where
drawing upon a variety of other reconstruction efforts may be helpful: They
suggest that certain technical matters arising in the course of a
reconstruction (e.g., the choice of a particular symmetry group) can
be dispatched with relative ease.

In special relativity and in thermodynamics, one builds up the theory
starting from postulates, the first of which has the character of a
guarantee. (Inertial observers Alice and Bob can come to agree on the
laws of physics; energy is conserved.) The second is a foil to the
first, frustrating it and generating a degree of dramatic
tension. (Alice and Bob cannot agree on a standard of rest, even by
measuring the speed of light; entropy is nondecreasing.)  Then comes a
statement of unattainability, which is derived in one case (massive
bodies cannot attain light speed) and assumed in the other (we cannot
cool all the way to absolute zero). We might also draw an analogy
between the clock postulate of special relativity, which lets us
analyze accelerated motion using momentarily co-moving inertial
frames, and the zeroth law of thermodynamics, which as it is applied
in practice is a statement about momentary equilibrium between systems
being a transitive condition. Might a similar story hold true for
quantum mechanics as well? In the view of quantum theory we are
developing here, the possibility of probability-1 predictions might be
considered a guarantee (certainty is allowed). The urgleichung is then
an axiom of frustration (certainty cannot be about hidden
variables). Perhaps the rejection of hidden variables, carefully
formulated in the urgleichung, will one day be recognized as the
Second Law of Quantum Mechanics.

\bibliographystyle{utphys}

\bibliography{mermin-bell-urgleichung}

\providecommand{\href}[2]{#2}\begingroup\raggedright\begin{thebibliography}{10}

\bibitem{Schroedinger:1935}
E.~Schr\"odinger, ``Discussion of probability relations between separated
  subsystems,'' \href{http://dx.doi.org/10.1017/S0305004100013554}{{\em Math.
  Proc. Cam. Phil. Soc.} {\bf 31} (1935) no.~4, 555--63}.

\bibitem{Barrett:2007}
J.~Barrett, ``Information processing in generalized probabilistic theories,''
  \href{http://dx.doi.org/10.1103/PhysRevA.75.032304}{{\em Phys. Rev. A} {\bf
  75} (2007) no.~3, 032304}, \href{http://arxiv.org/abs/quant-ph/0508211}{{\tt
  arXiv:quant-ph/0508211}}.

\bibitem{Barnum:2012}
H.~Barnum, J.~Barrett, M.~Leifer, and A.~Wilce, ``Teleportation in general
  probabilistic theories,'' {\em Proceedings of Symposia in Applied
  Mathematics} {\bf 71} (2012)  25--48,
  \href{http://arxiv.org/abs/0805.3553}{{\tt arXiv:0805.3553}}.

\bibitem{Werner:1989}
R.~F. Werner, ``Quantum states with {Einstein}--{Podolsky}--{Rosen}
  correlations admitting a hidden-variable model,''
  \href{http://dx.doi.org/10.1103/PhysRevA.40.4277}{{\em Phys. Rev. A} {\bf 40}
  (1989) no.~6, 4277}.

\bibitem{Gottesman:1999}
D.~Gottesman, ``The {H}eisenberg representation of quantum computers,'' in {\em
  Group22: Proceedings of the XXII International Colloquium on Group
  Theoretical Methods in Physics}, S.~P. Corney, R.~Delbourgo, and P.~D.
  Jarvis, eds.
\newblock International Press, 1999.
\newblock \href{http://arxiv.org/abs/quant-ph/9807006}{{\tt
  arXiv:quant-ph/9807006}}.

\bibitem{Spekkens:2007}
R.~W. Spekkens, ``Evidence for the epistemic view of quantum states: A toy
  theory,'' \href{http://dx.doi.org/10.1103/PhysRevA.75.032110}{{\em Phys. Rev.
  A} {\bf 75} no.~3, 032110}, \href{http://arxiv.org/abs/quant-ph/0401052}{{\tt
  arXiv:quant-ph/0401052}}.

\bibitem{Spekkens:2016b}
R.~W. Spekkens,
  \href{http://dx.doi.org/10.1007/978-94-017-7303-4_4}{``Quasi-quantization:
  Classical statistical theories with an epistemic restriction,''} in {\em
  Quantum Theory: Informational Foundations and Foils}, G.~Chiribella and R.~W.
  Spekkens, eds., pp.~83--135.
\newblock Springer, 2016.
\newblock \href{http://arxiv.org/abs/1409.5041}{{\tt arXiv:1409.5041}}.

\bibitem{Mermin:1990}
N.~D. Mermin, ``What's wrong with these elements of reality?,''
  \href{http://dx.doi.org/10.1063/1.2810588}{{\em Physics Today} {\bf 43}
  (1990) no.~6, 9}. Reprinted in \booktitle{{Why Quark Rhymes With Pork}}
  ({C}ambridge {U}niversity {P}ress, 2016), pp.\ 43--49.

\bibitem{Mermin:1990b}
N.~D. Mermin, ``Extreme quantum entanglement in a superposition of
  macroscopically distinct states,''
  \href{http://dx.doi.org/10.1103/PhysRevLett.65.1838}{{\em Phys. Rev. Lett.}
  {\bf 65} (1990) no.~15, 1838--40}.

\bibitem{Mermin:1993}
N.~D. Mermin, ``Hidden variables and the two theorems of {John} {Bell},''
  \href{http://dx.doi.org/10.1103/RevModPhys.65.803}{{\em Rev. Mod. Phys.} {\bf
  65} (1993) no.~3, 803--15}, \href{http://arxiv.org/abs/1802.10119}{{\tt
  arXiv:1802.10119}}.

\bibitem{Mermin:2016}
N.~D. Mermin, ``Erratum: {H}idden variables and the two theorems of {John}
  {Bell},'' \href{http://dx.doi.org/10.1103/RevModPhys.88.039902}{{\em Rev.
  Mod. Phys.} {\bf 88} (2016) no.~3, 039902}.

\bibitem{Hoggar:1981}
S.~G. Hoggar, \href{http://dx.doi.org/10.1007/978-1-4612-5648-9_14}{``Two
  quaternionic 4-polytopes,''} in {\em The Geometric Vein: The Coxeter
  Festschrift}, C.~Davis, B.~Gr\"unbaum, and F.~A. Sherk, eds.
\newblock Springer, 1981.

\bibitem{Hoggar:1998}
S.~G. Hoggar, ``64 lines from a quaternionic polytope,''
  \href{http://dx.doi.org/10.1023/A:1005009727232}{{\em Geometriae Dedicata}
  {\bf 69} (1998)  287--289}.

\bibitem{Szymusiak:2015}
A.~Szymusiak and W.~S{\l}omczy{\'{n}}ski, ``Informational power of the {H}oggar
  symmetric informationally complete positive operator-valued measure,''
  \href{http://dx.doi.org/10.1103/PhysRevA.94.012122}{{\em Phys. Rev. A} {\bf
  94} (2015)  012122}, \href{http://arxiv.org/abs/1512.01735}{{\tt
  arXiv:1512.01735}}.

\bibitem{Stacey:2016}
B.~C. Stacey, ``Sporadic {SIC}s and the normed division algebras,''
  \href{http://dx.doi.org/10.1007/s10701-017-0087-2}{{\em Found. Phys.} {\bf
  47} (2017)  1060--64}, \href{http://arxiv.org/abs/1605.01426}{{\tt
  arXiv:1605.01426}}.

\bibitem{Stacey:2016b}
B.~C. Stacey, ``Geometric and information-theoretic properties of the {H}oggar
  lines,'' \href{http://arxiv.org/abs/1609.03075}{{\tt arXiv:1609.03075}}.

\bibitem{Wootters:2009}
W.~K. Wootters, ``Symmetric informationally complete measurements: Can we make
  big ones out of small ones?,'' \href{https://pirsa.org/09120023}{{\tt
  PIRSA:09120023}}.

\bibitem{Fuchs:2013}
C.~A. Fuchs and R.~Schack, ``Quantum-{B}ayesian coherence,''
  \href{http://dx.doi.org/10.1103/RevModPhys.85.1693}{{\em Rev. Mod. Phys.}
  {\bf 85} (2013)  1693--1715}, \href{http://arxiv.org/abs/1301.3274}{{\tt
  arXiv:1301.3274}}.

\bibitem{Stacey:2016c}
B.~C. Stacey, ``{SIC-POVMs} and compatibility among quantum states,''
  \href{http://dx.doi.org/10.3390/math4020036}{{\em Mathematics} {\bf 4} (2016)
  no.~2, 36}, \href{http://arxiv.org/abs/1404.3774}{{\tt arXiv:1404.3774}}.

\bibitem{Fuchs:2016b}
C.~A. Fuchs and B.~C. Stacey,
  \href{http://dx.doi.org/10.1007/978-94-017-7303-4_9}{``Some negative remarks
  on operational approaches to quantum theory,''} in {\em Quantum Theory:
  Informational Foundations and Foils}, G.~Chiribella and R.~W. Spekkens, eds.,
  pp.~283--305.
\newblock Springer, 2016.
\newblock \href{http://arxiv.org/abs/1401.7254}{{\tt arXiv:1401.7254}}.

\bibitem{Fuchs:2016a}
C.~A. Fuchs and B.~C. Stacey, ``{QB}ism: Quantum theory as a hero's handbook,''
  \href{http://arxiv.org/abs/1612.07308}{{\tt arXiv:1612.07308}}.

\bibitem{Fuchs:2017b}
C.~A. Fuchs, ``Notwithstanding {B}ohr, the {R}easons for {QB}ism,'' {\em Mind
  and Matter} {\bf 15} (2017) no.~2, 245--300,
  \href{http://arxiv.org/abs/1705.03483}{{\tt arXiv:1705.03483}}.

\bibitem{DeBrota:2018}
J.~B. DeBrota, C.~A. Fuchs, and B.~C. Stacey, ``Symmetric informationally
  complete measurements identify the essential difference between classical and
  quantum,'' \href{http://arxiv.org/abs/1805.08721}{{\tt arXiv:1805.08721}}.

\bibitem{Appleby:2017b}
M.~Appleby, S.~Flammia, G.~McConnell, and J.~Yard, ``{SIC}s and algebraic
  number theory,'' \href{http://dx.doi.org/10.1007/s10701-017-0090-7}{{\em
  Found. Phys.} {\bf 47} (2017)  1042--59},
  \href{http://arxiv.org/abs/1701.05200}{{\tt arXiv:1701.05200}}.

\bibitem{Waldron:2018}
S.~Waldron, \href{http://dx.doi.org/10.1007/978-0-8176-4815-2}{{\em An
  Introduction to Finite Tight Frames}}.
\newblock Springer, 2018.
\newblock
  \url{https://www.math.auckland.ac.nz/~waldron/Preprints/Frame-book/frame-book.html}.

\bibitem{VonNeumann:1955}
J.~{von Neumann}, {\em Mathematical Foundations of Quantum Mechanics}.
\newblock Princeton University Press, 1955.

\bibitem{Werner:2014}
R.~F. Werner, ``Comment on {Maudlin}'s paper `{What} {Bell} did',''
  \href{http://dx.doi.org/10.1088/1751-8113/47/42/424011}{{\em J. Phys. A} {\bf
  47} (2014) no.~42, 424011}.

\bibitem{Stacey:2016d}
B.~C. Stacey, ``Von {Neumann} was not a {Quantum} {Bayesian},''
  \href{http://dx.doi.org/10.1098/rsta.2015.0235}{{\em Phil. Trans. Roy. Soc.
  A} {\bf 374} (2016)  20150235}, \href{http://arxiv.org/abs/1412.2409}{{\tt
  arXiv:1412.2409}}.

\bibitem{Mermin:2018}
N.~D. Mermin and R.~Schack, ``Homer nodded: von {Neumann's} surprising
  oversight,'' \href{http://arxiv.org/abs/1805.10311}{{\tt arXiv:1805.10311}}.

\bibitem{Fuchs:2014b}
C.~A. Fuchs, N.~D. Mermin, and R.~Schack, ``An introduction to {QBism} with an
  application to the locality of quantum mechanics,''
  \href{http://dx.doi.org/10.1119/1.4874855}{{\em Am. J. Phys.} {\bf 82} (2014)
  no.~8, 749--54}, \href{http://arxiv.org/abs/1311.5253}{{\tt
  arXiv:1311.5253}}.

\bibitem{Busch:2003}
P.~Busch, ``Quantum states and generalized observables: {A} simple proof of
  {Gleason's} theorem,''
  \href{http://dx.doi.org/10.1103/PhysRevLett.91.120403}{{\em Phys. Rev. Lett.}
  {\bf 91} (2003)  120403}, \href{http://arxiv.org/abs/quant-ph/9909073}{{\tt
  arXiv:quant-ph/9909073}}.

\bibitem{Caves:2004}
C.~M. Caves, C.~A. Fuchs, K.~K. Manne, and J.~M. Renes, ``Gleason-type
  derivations of the quantum probability rule for generalized measurements,''
  \href{http://dx.doi.org/10.1023/B:FOOP.0000019581.00318.a5}{{\em Found.
  Phys.} {\bf 34} (2004)  193--209},
  \href{http://arxiv.org/abs/quant-ph/0306179}{{\tt arXiv:quant-ph/0306179}}.

\bibitem{Caves:2002c}
C.~M. Caves, C.~A. Fuchs, and R.~Schack, ``Unknown quantum states: The quantum
  de {F}inetti representation,''
  \href{http://dx.doi.org/10.1063/1.1494475}{{\em J. Math. Phys.} {\bf 43}
  (2002) no.~9, 4537--59}, \href{http://arxiv.org/abs/quant-ph/0104088}{{\tt
  arXiv:quant-ph/0104088}}.

\bibitem{Appleby:2007a}
D.~M. Appleby, ``Symmetric informationally complete measurements of arbitrary
  rank,'' \href{http://dx.doi.org/10.1134/S0030400X07090111}{{\em Opt. Spect.}
  {\bf 103} (2007)  416--428},
  \href{http://arxiv.org/abs/quant-ph/0611260}{{\tt arXiv:quant-ph/0611260}}.

\bibitem{Zauner:1999}
G.~Zauner, \href{http://dx.doi.org/10.1142/S0219749911006776}{{\em
  Quantendesigns. {G}rundz{\"u}ge einer nichtkommutativen {D}esign\-theorie}}.
\newblock PhD thesis, University of Vienna, 1999.
\newblock \url{http://www.gerhardzauner.at/qdmye.html}.
\newblock Published in English translation: G.~Zauner, ``Quantum designs:
  foundations of a noncommutative design theory,'' \emph{Int. J. Quantum
  Inf.}~\textbf{9} (2011) 445--508.

\bibitem{Renes:2004}
J.~M. Renes, R.~Blume-Kohout, A.~J. Scott, and C.~M. Caves, ``Symmetric
  informationally complete quantum measurements,''
  \href{http://dx.doi.org/10.1063/1.1737053}{{\em J. Math. Phys.} {\bf 45}
  (2004)  2171--2180}.

\bibitem{Scott:2010a}
A.~J. Scott and M.~Grassl, ``Symmetric informationally complete
  positive-operator-valued measures: A new computer study,''
  \href{http://dx.doi.org/10.1063/1.3374022}{{\em J. Math. Phys.} {\bf 51}
  (2010)  042203}.

\bibitem{Fuchs:2017a}
C.~A. Fuchs, M.~C. Hoang, and B.~C. Stacey, ``The {SIC} question: {History} and
  state of play,'' \href{http://dx.doi.org/10.3390/axioms6030021}{{\em Axioms}
  {\bf 6} (2017)  21}, \href{http://arxiv.org/abs/1703.07901}{{\tt
  arXiv:1703.07901}}.

\bibitem{Feynman:1987}
R.~P. Feynman, ``Negative probability,'' in {\em Quantum Implications: Essays
  in Honour of {D}avid {B}ohm}, pp.~235--48.
\newblock Routledge, 1987.
\newblock \url{http://cds.cern.ch/record/154856}.

\bibitem{Ferrie:2008}
C.~Ferrie and J.~Emerson, ``Frame representations of quantum mechanics and the
  necessity of negativity in quasi-probability representations,''
  \href{http://dx.doi.org/10.1088/1751-8113/41/35/352001}{{\em J. Phys. A} {\bf
  41} (2008)  352001}.

\bibitem{Ferrie:2009}
C.~Ferrie and J.~Emerson, ``Framed {H}ilbert space: hanging the
  quasi-probability pictures of quantum theory,''
  \href{http://dx.doi.org/10.1088/1367-2630/11/6/063040}{{\em New J. Phys.}
  {\bf 11} (2009)  063040}.

\bibitem{Zhu:2016a}
H.~Zhu, ``Quasiprobability representations of quantum mechanics with minimal
  negativity,'' \href{http://dx.doi.org/10.1103/PhysRevLett.117.120404}{{\em
  Phys. Rev. Lett.} {\bf 117} (2016) no.~12, 120404},
  \href{http://arxiv.org/abs/1604.06974}{{\tt arXiv:1604.06974}}.

\bibitem{DeBrota:2017}
J.~B. DeBrota and C.~A. Fuchs, ``Negativity bounds for {W}eyl--{H}eisenberg
  quasiprobability representations,'' {\em Found. Phys.} {\bf 47} (2017)
  1009--30, \href{http://arxiv.org/abs/1703.08272}{{\tt arXiv:1703.08272}}.

\bibitem{Gieser:2005}
S.~Gieser, {\em The Innermost Kernel: Depth Psychology and Quantum Physics.
  Wolfgang Pauli's Dialogue with C.\ G.\ Jung}.
\newblock Springer-Verlag, 2005.

\bibitem{Atmanspacher:2009}
H.~Atmanspacher and H.~Primas, eds., {\em Recasting Reality: Wolfgang Pauli's
  Philosphical Ideas and Contemporary Science}.
\newblock Springer-Verlag, 2009.

\bibitem{Appleby:2016b}
M.~Appleby, C.~A. Fuchs, B.~C. Stacey, and H.~Zhu, ``Introducing the {Qplex}: A
  novel arena for quantum theory,'' {\em Eur. Phys. J. D} {\bf 71} (2016)  197,
  \href{http://arxiv.org/abs/1612.03234}{{\tt arXiv:1612.03234}}.

\bibitem{Jedwab:2015}
J.~Jedwab and A.~Wiebe,
  \href{http://dx.doi.org/10.1007/978-3-319-17729-8_13}{``A simple construction
  of complex equiangular lines,''} in {\em Algebraic Design Theory and Hadamard
  Matrices}, pp.~159--169.
\newblock Springer, 2015.
\newblock \href{http://arxiv.org/abs/1408.2492}{{\tt arXiv:1408.2492}}.

\bibitem{Zhu:2012}
H.~Zhu, {\em Quantum state estimation and symmetric informationally complete
  {POM}s}.
\newblock PhD thesis, National University of Singapore, 2012.
\newblock \url{http://scholarbank.nus.edu.sg/handle/10635/35247}.

\bibitem{Spekkens:2016}
R.~W. Spekkens, ``Reassessing claims of nonclassicality for quantum
  interference phenomena,'' \href{https://pirsa.org/16060102}{{\tt
  PIRSA:16060102}}.

\bibitem{Pusey:2015}
M.~F. Pusey and M.~S. Leifer, ``Logical pre- and post-selection paradoxes are
  proofs of contextuality,'' \href{http://dx.doi.org/10.4204/EPTCS.195.22}{{\em
  EPTCS} {\bf 195} (2015)  295--306},
  \href{http://arxiv.org/abs/1506.07850}{{\tt arXiv:1506.07850}}.

\bibitem{Einstein:1935}
A.~Einstein, B.~Podolsky, and N.~Rosen, ``Can quantum-mechanical description of
  physical reality be considered complete?,''
  \href{http://dx.doi.org/10.1103/PhysRev.47.777}{{\em Phys. Rev.} {\bf 47}
  (1935)  777--80}.

\bibitem{Mermin:1985}
N.~D. Mermin, ``Is the moon there when nobody looks? {R}eality and the quantum
  theory,'' \href{http://dx.doi.org/10.1063/1.880968}{{\em Physics Today} {\bf
  38} (1985) no.~4, 38--47}.

\bibitem{Spekkens:2014}
R.~W. Spekkens, ``The status of determinism in proofs of the impossibility of a
  noncontextual model of quantum theory,''
  \href{http://dx.doi.org/10.1007/s10701-014-9833-x}{{\em Found. Phys.} {\bf
  44} (2014) no.~11, 1125--55}, \href{http://arxiv.org/abs/1312.3667}{{\tt
  arXiv:1312.3667}}.

\bibitem{Spekkens:2005}
R.~W. Spekkens, ``Contextuality for preparations, transformations, and unsharp
  measurements,'' \href{http://dx.doi.org/10.1103/PhysRevA.71.052108}{{\em
  Phys. Rev. A} {\bf 71} (2005) no.~5, 052108},
  \href{http://arxiv.org/abs/quant-ph/0406166}{{\tt arXiv:quant-ph/0406166}}.

\bibitem{Bravyi:2005}
S.~Bravyi and A.~Kitaev, ``Universal quantum computation with ideal {C}lifford
  gates and noisy ancillas,''
  \href{http://dx.doi.org/10.1103/PhysRevA.71.022316}{{\em Phys. Rev. A} {\bf
  71} (2005)  022316}, \href{http://arxiv.org/abs/quant-ph/0403025}{{\tt
  arXiv:quant-ph/0403025}}.

\bibitem{Andersson:2017}
O.~Andersson, P.~Badzl\c{a}g, I.~Bengtsson, I.~Dumitru, and A.~Cabello,
  ``Self-testing properties of {G}isin's elegant {B}ell inequality,''
  \href{http://dx.doi.org/10.1103/PhysRevA.96.032119}{{\em Phys. Rev. A} {\bf
  96} (2017)  032119}, \href{http://arxiv.org/abs/1706.02130}{{\tt
  arXiv:1706.02130}}.

\bibitem{Bengtsson:2012}
I.~Bengtsson, K.~Blanchfield, and A.~Cabello, ``A {K}ochen-{S}pecker inequality
  from a {SIC},'' \href{http://dx.doi.org/10.1016/j.physleta.2011.12.011}{{\em
  Phys. Lett. A} {\bf 376} (2012)  374--376},
  \href{http://arxiv.org/abs/1109.6514}{{\tt arXiv:1109.6514}}.

\bibitem{Barnum:2014}
H.~Barnum, M.~P. M\"uller, and C.~Ududec, ``Higher-order interference and
  single-system postulates characterizing quantum theory,''
  \href{http://dx.doi.org/10.1088/1367-2630/16/12/123029}{{\em New J. Phys.}
  {\bf 16} (2014)  123029}, \href{http://arxiv.org/abs/1403.4147}{{\tt
  arXiv:1403.4147}}.

\bibitem{Barnum:2015}
H.~Barnum, M.~Graydon, and A.~Wilce, ``Some nearly quantum theories,'' {\em
  EPTCS} {\bf 195} (2015)  59--70, \href{http://arxiv.org/abs/1507.06278}{{\tt
  arXiv:1507.06278}}.

\bibitem{Wilce:2016}
A.~Wilce, ``A royal road to quantum theory (or thereabouts),''
  \href{http://arxiv.org/abs/1606.09306}{{\tt arXiv:1606.09306}}.

\bibitem{Krumm:2016}
M.~Krumm, H.~Barnum, J.~Barrett, and M.~P. M\"uller, ``Thermodynamics and the
  structure of quantum theory,'' \href{http://arxiv.org/abs/1608.04461}{{\tt
  arXiv:1608.04461}}.

\bibitem{Cabello:2018}
A.~Cabello, ``A simple explanation of {Born}'s rule,''
  \href{http://arxiv.org/abs/1801.06347}{{\tt arXiv:1801.06347}}.

\bibitem{VanDeWetering:2018}
J.~{van de Wetering}, ``Sequential measurement characterizes quantum theory,''
  \href{http://arxiv.org/abs/1803.11139}{{\tt arXiv:1803.11139}}.

\bibitem{Chiribella:2011}
G.~Chiribella, G.~M. D'Ariano, and P.~Perinotti, ``Informational derivation of
  quantum theory,'' \href{http://dx.doi.org/10.1103/PhysRevA.84.012311}{{\em
  Phys. Rev. A} {\bf 84} (2011) no.~1, 012311},
  \href{http://arxiv.org/abs/1011.6451}{{\tt arXiv:1011.6451}}.

\bibitem{Disilvestro:2017}
L.~Disilvestro and D.~Markham, ``Quantum protocols within {Spekkens'} toy
  model,'' \href{http://dx.doi.org/10.1103/PhysRevA.95.052324}{{\em Phys. Rev.
  A} {\bf 95} (2017) no.~5, 052324},
  \href{http://arxiv.org/abs/1608.09012}{{\tt arXiv:1608.09012}}.

\end{thebibliography}\endgroup

\end{document}